# Optimizing Data Cube Visualization for Web Applications

Performance and User-Friendly Data Aggregation

Daniel Szelogowski

UW - Whitewater
Computer Science Master's Program
Whitewater, WI
szelogowdj19@uww.edu

**Abstract.** *Current open-source applications which allow for cross-platform data visualization of OLAP cubes feature issues of high overhead and inconsistency due to data oversimplification. To improve upon this issue, there is a need to cut down the number of pipelines that the data must travel between for these aggregation operations and create a single, unified application which performs efficiently without sacrificing data, and allows for ease of usability and extension.*

## 1. INTRODUCTION

With open source, non-commercial applications becoming both more popular and widely accepted, data manipulation software is still lacking options and support for development. Particularly in **Online Analytical Processing (OLAP)** systems, which allow users to easily aggregate and manipulate complex data from multiple database systems simultaneously and analyze data from multiple extracted viewpoints [9], there is a lack of non-commercial grade software for performing these database operations — especially for visualizing the data, a task becoming ever-increasingly popular and necessary on varying platforms such as web applications which typically allow for the most cross-platform support. As WebAssembly, the new "fourth language of the web" increasingly matures, web applications have become much more practical due to their potential for near-native runtimes and extensibility of natively compiled code.

This paper seeks to discuss current web applications and their implementations for visualizing and aggregating **OLAP Cubes** (or **Data Cubes**), the primary data structure where databases are divided into to allow ease of analysis and visualization, issues across these applications, and potential solutions to create a more runtime-efficient, user-friendly application.

## 2. RELATED WORK

In previous work [1], I discussed various implementations of open source, web application-ready libraries/implementations of OLAP cube viewers, particularly the Python library Atoti for Jupyter Notebooks and the **HTML5 application** CubesViewer. The lack of implementations lead me to these two, both of which have two issues I sought to alleviate:

- **Runtime**
- **Data Oversimplification**

In particular, I focused on the issues of their implementations — Atoti, being written in Python, suffers from the performance hit of being an interpreted programming language (especially due to the **Global Interpreter Lock** and lacking **concurrency-related** features) and has no native support for **high-precision floating-point numbers**, requiring external libraries such as NumPy [2] which then must be passed to a Java interoperability system further slowing the process, and CubesViewer as an HTML5 application suffers especially from the backend being written in JavaScript (JS), which is also dynamically typed, single-threaded, and also lacks support for 64-bit integers and high-precision floating-point numbers [3]. Other applications

## 2.1 Cubes Server

Cubes Server is an open-source OLAP framework and HTTP server written in Python [4]. The application features an OLAP Cube aggregation library and a Slicer HTTP server which hosts the data cube and performs aggregation functions on the data to return the result. Data must first be prepared using a csv-to-sqlite file conversion script and have aggregation features selected through hard-coding before being served using Slicer.

## 2.2 CubesViewer

CubesViewer is an HTML5 application featuring responsive data visualization and aggregation used particularly for visualizing data cubes. In order to perform the OLAP Cube operations, CubesViewer uses JavaScript functions to parse the operation chosen by the user and sends a request to a "Cubes Server" which hosts the data and performs the aggregation operations and returns the computed data table. CubesViewer features five primary operations:

- **Display Fact Table** - Display the full (unaggregated) data table
- **Aggregate** - The simplest operation; sums all data of a particular column and returns the sum and record count (or number of items used in the summary)
- **Drill-down** - Add an additional variable (column) to the data view
- **Filter** - Filter a drilled-down column by a particular value
- **View (Plot)** - Generate a plot based on user-selected independent and dependent variables (columns)

CubesViewer performs these operations well, albeit somewhat slow due to its double-backend system, but also lacks ease of usability for the common-user. The only way to view data is to prepare a Python script to select the aggregation variables and code them in by their operation usage, convert the data table into sqlite, and create an ini file to be served on an HTTP server using a Slicer Server [5]. This is one of the key problems this paper seeks to resolve regarding user-friendliness, as well as the issue of CubesViewer needing to perform computations in both the JavaScript and Python backend systems. CubesViewer also features both a client and server form, however the server form simply extends the client application to add features such as view saving and sharing; this is also written in python using Django to serve the web pages [6].

## 3. APPROACH

Due to the unfortunate complexity of these programs, the easiest method of reducing this was to rewrite the functionality of these programs into one sole application which would serve the purpose of all three others. While an implementation in JavaScript would be possible to reduce our application solely to HTML5 + JS, this would not only perform poorly, but would be an extremely challenging task computationally and programatically due to the lack of data science libraries for the language.

In its simplest form, we begin with a simple console-based application or script that can read any Cube dataset and perform all four aggregation operations and allow the user to create varying visualizations of the data. After this, we can modify the application to perform additional functions such as serving computations across HTTP headers, thereby reducing the system to a single computational backend, and/or modifying the application to utilize WebAssembly to create an all-in-one system that requires no additional server or backend and can be packaged as a single independent application, such as using **Blazor WebAssembly** in C# [10] which would also allow for the utilization of concurrency to reduce program runtime.

## 4. IMPLEMENTATION

To begin my approach, I created a Python script to condense and recreate the functionality of the three programs (**CubesViewer Client** and **Server**, **Cube Server**) in one single application — this reduction allowed for a massive compression and slight performance boost from a reduced library

requirement — becoming a 200-line Python script, albeit lacking a web interface (all implementations can be found in [8]).

## 4.1 Python

The Python script performs the four aggregation functions described in *Section 2.2*, as well as allowing the user to change the initial aggregated feature (column), add/remove/clear filters or drill-down aggregates at any time, view both the fact table (original data) and current aggregated table, or generate a graph as one of the following:
- Bar
- Line (with or without a marker)
- Scatter
- Pie

Before generating the graph, the user must choose the features to be used for the horizontal and vertical axes which are then automatically labeled by their corresponding feature in the plot. The plot is generated through **Matplotlib's PyPlot** library, allowing for user interaction and figure saving of the user's view — this alone removes the case for CubesViewer Server. Optionally, additional code is provided which converts the returned figure into a base64 string and generates an HTML image tag pre-sourced to our generated plot. Table 1 shows an initial aggregate table, and a drilled-down table can be seen in Table 2. These generated data tables are provided to match the output of **CubesViewer Studio** (client).

| Summary | Sum of All "Amount (US$-Millions)" | Record Count |
|---|---|---|
| Summary | 558430000.0 | 31000 |

**Table 1: Initial Aggregate Table**

| Category | Subcategory Code | Fiscal Year | Amount (US$-Millions) |
|---|---|---|---|
| Assets | dfb | 2010 | 1581 |
| Assets | dfb | 2009 | 2380 |
| ... | ... | ... | ... |
| Equity | oe | 2009 | -1683 |

**Table 2: Example table after performing a drill-down on the features "Category", "Subcategory Code" and "Fiscal Year"**

As well, the user also has the ability to load any dataset with no necessary modifications (such as Cube Server's preparation and aggregate-prepare scripts and conversion to sqlite) so long as it meets the following requirements:

1. The data file must be a CSV (although this may be modified in the future)
2. The first row of the data file must be the column headers

The script will automatically parse the headers and store them for use as potential aggregate features, eliminating the hard-coding requirement for the user.

## 4.2 C# - The Path to Concurrent WebAssembly

Implementing the same functionality in Golang with the current features of the language, due to the lack of dynamic typing necessitating the use of the generic "interface{}" type which is error-prone, so the next ideal candidate language which features strong concurrency and compilation to WebAssembly when desired was C#.

Implementation for C# is a much greater task than in Python, due to Python's stronger support for dynamic typing and vast library of data science related libraries and tools; while NumPy binding and clone libraries exist for C#, the "DataTable" class was the closest flexible option offered by the language, providing similar features to NumPy's ND-Array type with lower overhead than C#'s built-in jagged and multidimensional array types. For plotting, using Windows Forms and the library "**OxyPlot**" was the closest candidate to providing the same functionality as Python's Matplotlib — Figure 1 shows an example comparison between the two. Notably, the data output by Figure 1(a) is displayed sorted, which appears to be performed by OxyPlot during figure generation, whereas in Figure 1(b), Matplotlib displays the data in order of the cube. This is easily solvable (shown in Figure 1(c)) by attaching Numpy's sort/argsort functionality, but will increase runtime during evaluation.

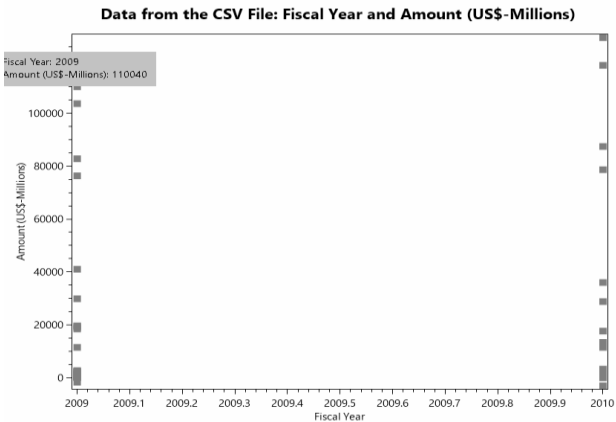

**Figure 1(a): OxyPlot output (C#)**

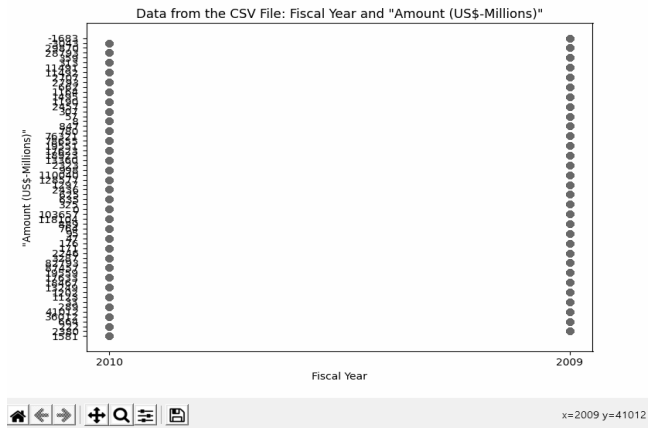

**Figure 1(b): Matplotlib output (Python) before sort modification to match OxyPlot**

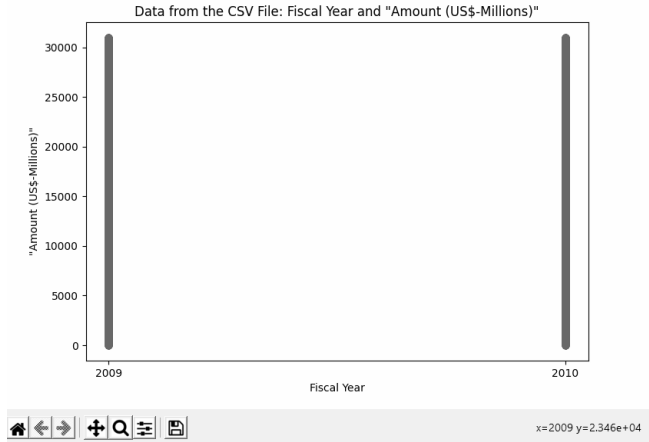

**Figure 1(c): Matplotlib output (Python) after sort modification to match OxyPlot**

## 5. EVALUATION

To demonstrate the potential for C# over Python, we have two major focal points:
- Show that C# runs faster due to the Common Language Runtime performing near-natively
- Show that concurrency does increase performance of computationally-expensive functions wherever possible

Primarily, we want to evaluate the performance of the four aggregation operations and the graph generation, of which will be measured in seconds (s). In Python and the C# applications, we will time these functions using a 31,000-record dataset (excluding the headers) of 7 features/columns (the dataset can be found in [8]). As well, the C# application will be broken into two copies: one which performs most operations **serially** (iteratively), and one which utilizes **concurrency** (distributed across multiple CPU threads/cores).

Due to the current lacking library support for Python-like operations such as list slices, however, many operations in C# are forcibly required to remain serial to prevent data loss or program errors, as concurrent looping operations have no guarantee of being performed in-order due to their distributed nature, though this is potentially reducible through additional research on extension methods for the DataTable objects.

To demonstrate, the following steps will be performed and measured:
1. Display fact table
2. Display initial aggregate table
3. Drill-down 3 features (columns 1, 3, 5), display aggregated table after each
4. Apply filter (column 5, 2009), display aggregated table
5. Remove filter, display aggregated table
6. Generate scatter plot on column 5 (x) and column 6 (y) data (same as Figure 1)

The results of these experiments are shown in Table 3 and Figure 2.

|  | **Python** | **C#** | **C#-Concurrent** |
|---|---|---|---|
| **Step 1** | 72.5291 | 70.9802 | 65.1369 |
| **Step 2** | 0.0617 | 0.0157 | 0.0092 |
| **Step 3** | 78.6145 | 87.2508 | 70.9069 |
| **Step 4** | 15.1480 | 14.2058 | 8.0208 |
| **Step 5** | 28.6886 | 14.2974 | 8.1545 |
| **Step 6** | 3.8543 | 2.1755 | 1.7510 |

**Table 3: Performance evaluation of all three application forms (measured in seconds), average across 100 trials**

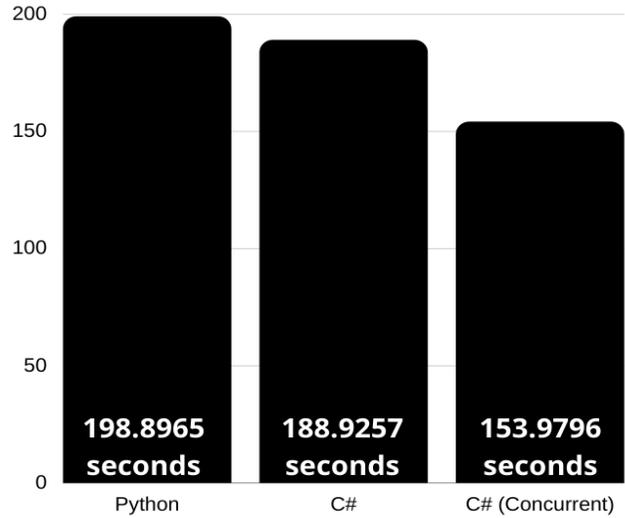

**Figure 2(a): Average runtime of each application's exam (sum of all 6 steps across 100 trials)**

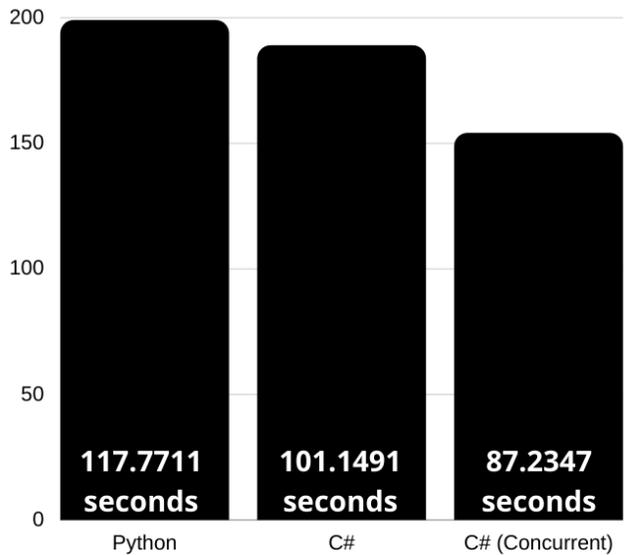

**Figure 2(b): Average runtime of each application's aggregation functions (sum of all 4 aggregation operations and view function across 100 trials)**

## 6. DISCUSSION

In this paper, we discussed and implemented a system which utilizes a single-backend calculation framework and evaluated its performance between

Python and C#, which can be compiled to WebAssembly to perform at near-native speeds and introduce concurrency for increased optimization.

Future work can be done to extend the existing application scripts to finalize the approach of a single, condensed package, but this will be a challenge — in my own attempts, there were a few major issues I ran into:

**In Golang:**
- No support for **dynamic types**, generics are not yet fully implemented [7]. The "interface{}" type is not flexible enough to replace these features
- Compilation to **WebAssembly** is very new and still challenging to implement without error as opposed to C#; however, Golang WebAssembly does perform near-native speeds when correct [1]
- Lacking libraries for data science (no strong NumPy or Matplotlib equivalent)

**In Python:**
- Using **Flask** to receive tasks and serve computed data works well, but script-global variables are non-existent in Flask apps, requiring the use of session storage at best
- Session storage, however, does not work well with single-page applications, as the HTTP header which sent the "GET" message will differ from the page receiving the message, making our stored data inaccessible.

**In C#:**
- Lacking library support for plotting and JavaScript interop without using ASP.NET
- In **Blazor (WebAssembly)**, some functions behave differently or create absurd conflicts and errors
- Lacking standard-library features — DataTables work well, but only for rows of data. Column computations and modifications lack support and must be hand-written each time
- Variable storage works well in Blazor, but working with locally-uploaded files without posting to a server is very challenging

**In HTML5+JS:**
- JavaScript's many language limitations, such as needing to use **Ajax** to perform HTTP requests without refreshing the page
- Dynamically updating HTML for large computations through JS after receiving computed data from an HTTP response is both difficult and creates high overhead

Many of these issues are solvable through a growing community of data scientists working in these various languages implementing new libraries for these operations, or through updates to the language or compiler itself. As a result, most applications used to perform these tasks are written using a multi-language interop system.

## 7. CONCLUSION

This paper focused on compressing the features of Cube Server and CubesViewer Studio and Server into a single application with the goal of minimizing overhead to increase the speed of data manipulation operations, as well as enabling a more user-friendly means of utilizing and interacting with data. Through tests of serial versus concurrent operations and interpreted versus compiled languages such as C# (which features just-in-time compilation to MSIL) which can be compiled to WebAssembly, we found that utilizing C# over Python not only resulted in a reduced runtime, but applying concurrency also greatly increased performance.

Future work may be done to implement this system in C# Blazor WebAssembly as support and libraries for Blazor grow and C# continues to develop as a language, or utilizing Go once dynamic or generic typing are fully implemented. An HTTP service is also very possible, but a very difficult task.